\documentclass[letterpaper]{article}
\usepackage{aaai}
\usepackage{times}
\usepackage{helvet}
\usepackage{courier}
\usepackage{graphicx}
\usepackage{amsmath}
\usepackage{amsthm}
\usepackage{flushend}
\usepackage{multirow}
\usepackage{algorithm}
\usepackage{algorithmic}
\begin{document}
\title{Epidemic Intelligence for the Crowd, by the Crowd\footnotemark \\
{\small $[$Full Version$]$}
}

\author{Ernesto Diaz-Aviles$^1$, Avar\'{e} Stewart$^1$, Edward Velasco$^2$,\\ 
\textbf{Kerstin Denecke}$^1$, \and \textbf{Wolfgang Nejdl}$^1$  \\ \\
$^1$L3S Research Center / University of Hannover. Hannover, Germany. \\
\texttt{\{diaz, stewart, denecke, nejdl\}@L3S.de}\\
$^2$Robert Koch Institute. Berlin, Germany.\\
\texttt{VelascoE@rki.de}
}
\copyrighttext{*A short version of this work has been accepted for publication at the International AAAI Conference on Weblogs and Social Media (ICWSM 2012).}

\maketitle

\begin{abstract}
\begin{quote}
Tracking Twitter for public health has shown great potential. However, most recent work has been focused on correlating Twitter messages to influenza rates, a disease that exhibits a marked seasonal pattern. In the presence of sudden outbreaks, how can social media streams be used to strengthen surveillance capacity? In May 2011, Germany reported an outbreak of \textit{Enterohemorrhagic Escherichia coli} (EHEC). It was one of the largest described outbreaks of EHEC/HUS worldwide and the largest in Germany. In this work, we study the \textit{crowd}'s behavior in Twitter during the outbreak. In particular, we report how tracking Twitter helped to detect key user messages that triggered signal detection alarms before MedISys and other well established early warning systems. We also introduce a personalized learning to rank approach that exploits the relationships discovered by: (i) latent semantic topics computed using Latent Dirichlet Allocation (LDA), and (ii) observing the social tagging behavior in Twitter, to rank tweets for epidemic intelligence. Our results provide the grounds for new public health research based on social media.
\end{quote}
\end{abstract}

%\section{Epidemic Intelligence Based on Twitter}
\section{Epidemic Intelligence Based on Twitter}
\label{sec:Intro_EpidemicIntelligenceBasedOnTwitter}
In May 2011, an outbreak of enterohaemorrhagic \textit{Escherichia coli} (EHEC) occurred in northern Germany. It was one of the largest described outbreaks of EHEC/HUS worldwide and the largest in Germany \cite{Frank_Ecoli:doi:10.1056/NEJMoa1106483}. 

Day 1: May 19, 2011, the Robert Koch Institute (RKI), Germany's Federal Public Health Authority, was invited by the Health and Consumer Protection Agency in Hamburg to assist in the  investigation of three cases of Hemolytic-uremic syndrome (HUS), a life-threatening illness caused by EHEC. Day 2: May 20, alarmed by the type of persons affected and the rapid spread of EHEC, an investigation was initiated by RKI, involving all levels of public-health and food-safety authorities to identify the cause of the outbreak, and to prevent further cases of disease. On day 5: May 23, RKI asked \textit{all} health departments to expedite procedures, by immediately forwarding all case reports of suspected or confirmed EHEC/HUS, to the Federal Public Health Authority, relying directly on the diagnoses of notifying clinicians \cite{Frank_Ecoli:doi:10.1056/NEJMoa1106483,RKI2011:EHEC_MayJune}.

Based on this five-day timeline of EHEC/HUS 2011 outbreak in Germany, one can see that public health officials are faced with new challenges for outbreak alert and response. This is due to the continuous emergence of infectious diseases and their contributing factors such as demographic change, or globalization. Early reaction is necessary, but often communication and information flow through traditional channels is slow. \textit{Can additional sources of information, such as social media streams, provide complements to the traditional epidemic intelligence mechanisms?}

Epidemic Intelligence (EI) encompasses activities related to early warning functions, signal assessments and outbreak investigation. Only the early detection of disease activity, followed by a rapid response, can reduce the impact of epidemics. Recently, modern disease surveillance systems have started to also monitor social media streams, with the objective of improving their timeliness to detect disease outbreaks, and producing warnings against potential public health threats (e.g., \cite{Corley:2010:Int-J-Environ-Res-Public-Health:20616993}). The real-time nature of Twitter makes it even more attractive for public health surveillance.

Recent works have shown the potential of using Twitter for public health. These works have either focused on: the text classification and filtering of tweets~\cite{2012_IHI_Mustafa,Sriram2010}; or finding predictors for diseases that exhibit a seasonal pattern (i.e., influenza-like illnesses) by correlating selected keywords with official influenza statistics and rates~\cite{Culotta:2010,Lampos2010,Signorini2011}. Still others have focused on mining Twitter content for topic ~\cite{Paul_Dredze_2011,DBLP:conf/icwsm/PaulD11} or sentiment analysis~\cite{chew-2009}. Furthermore, these existing approaches have all focused on countries where the tweet density is known to be high (e.g., the UK, or U.S.). 

In this paper, we seek to address the issues that can help deliver a public health surveillance system based on Twitter, by taking into account two important stages in epidemic intelligence: \textit{Early Outbreak Detection} and \textit{Outbreak Analysis and Control}, and take up the following questions: 

\begin{enumerate}
 \item \textit{Early Outbreak Detection}: Is it possible, by only using Twitter, to find early cases of an outbreak, before well established systems?
 \item \textit{Outbreak Analysis and Control}: Is it possible to use Twitter to understand the potential causes of contamination and spread? and How can we provide support for public health official to analyze and assess the risk based on the available social media information? 
\end{enumerate}

In contrast, to the aforementioned studies, ours focuses on a sudden outbreak of a disease that does not involve any seasonal pattern. Moreover, our work shows the potential of Twitter in countries where the tweet density is significantly lower, such as Germany. The contributions of this paper are summarized as follows:

\begin{itemize}
\item We provide an example of the application of standard surveillance algorithms on Twitter data collected in real-time during a major outbreak of EHEC/HUS in Germany, and provide insights showing the potential of Twitter for early warning.

\item For outbreak analysis and control, many studies have been made for systems that return documents in response to a query, little effort has been devoted to exploiting learning to rank in a personalized setting, specially in the domain of epidemic intelligence. This paper presents an innovative personalized ranking approach that offers decision makers the most relevant and attractive tweets for risk assessment, by exploiting latent topics and social hash-tagging behavior in Twitter.
\end{itemize}

The rest of the paper is organized as follows: In Section \ref{sec:earlyWarning}, we show how an early warning based on Twitter is possible, we present the data collection used in our experiments and analysis, and the standard biosurveillance methods applied. In Section~\ref{sec:analysisAndControl}, we introduce a personalized learning to rank approach, based on Twitter, to support the task of analysis and control in the presence of a sudden outbreak. Related works are discussed in Section~\ref{sec:relwork}. Finally, in Section~\ref{sec:conclusionAndFutureDirection}, we summarize our findings, point to future directions, and conclude the paper.

\section{Twitter for Early Warning}
\label{sec:earlyWarning}
The continuous emergence of infectious diseases and their contributing factors impose new challenges to public health officials. Early reaction is necessary, but often communication and information flow through traditional channels is slow. Additional sources of information, such as social media streams, provide complements to the traditional reporting mechanisms. 

\begin{figure*}[ht!]
\centering
\includegraphics[height=0.35\textheight]{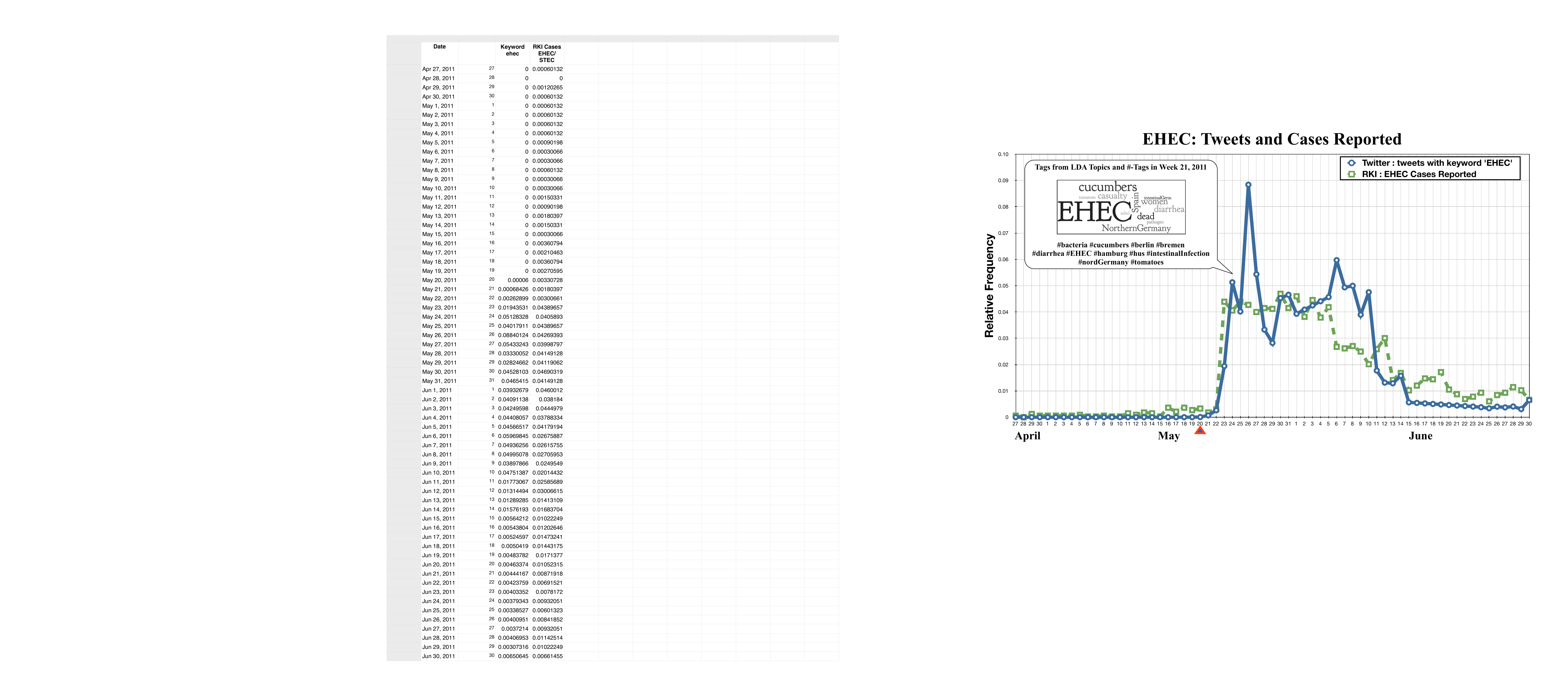}
\caption{\bf Relative frequency of cases reported to RKI and the number of tweets mentioning the name of the disease: \textit{EHEC}. The Pearson correlation coefficient is 0.864. Monitoring Twitter allowed us to generate the first signal on Friday, May 20th, 2011, using standard biosurveillance methods, before well established early warning systems (triangle on the time axis).}
\label{fig:EHEC_tweets_and_rki}
\vspace{-10pt}
\end{figure*}

For example, if we observe Figure \ref{fig:EHEC_tweets_and_rki} , we can see two plots, one of them corresponds to the relative frequency of EHEC cases as reported by RKI~\cite{RKI2011:EHEC_MayJune}, and the other to the relative frequency of mentions of the keyword ``EHEC" in the tweets collected during the months of May and June 2011. We can appreciate the high correlation of the curves, which corresponds to a Pearson correlation coefficient of 0.864. We can also observe the \textit{inertia} of the crowd that continued tweeting about the outbreak, even though the number of cases were already declining (e.g., June 5 to 11). 

Twitter has shown potential as a source of information for public health event monitoring (e.g.,~\cite{DBLP:conf/icwsm/PaulD11,2012_IHI_Mustafa}), but could it be possible to generate an early warning signal before well established systems by only tracking Twitter? 

In this section, we have a closer look to the time period of the EHEC/HUS outbreak in Germany, and address this question.

\subsection{Data Collection}
\label{sec:dataCollection}
We incrementally collected tweets using Twitter's API,  currently we monitor over 500 diseases and symptoms, which include ``EHEC". One of the challenges we face collecting data from Twitter, besides the API restrictions, is the level of \textit{noise} with respect to medical domain content. Straightforward techniques relying on regular expressions, even though they exhibit high recall, are difficult to maintain and prone to high false positive rates. For example, consider the following two tweets collected by a combination of regular expressions, and a dictionary of diseases that includes the medical conditions \textit{EHEC} and \textit{fever}:

\begin{enumerate}
	\item \textit{RKI warns against north German vegetables: Experts looking feverishly \textbf{EHEC} source http://bit.ly/itGpJx}
	\item \textit{I've definitely Bieber-\textbf{fever}. There's no doubt. but who hasn't got bieber \textbf{fever}? @justinbieber is soo damn rawwwr}
\end{enumerate}

Tweet number one is of obvious importance for epidemic intelligence, but number two is not. 

Instead of simple keyword matching to filter out irrelevant tweets, our data collection strategy includes text classification methods and a multi-level filtering based on supervised learning, following the approach of Stewart et al.~\cite{Stewart:2011}.

Table \ref{tab:dataCollectionStats} summarizes the data collected related to the outbreak that was used in our analysis.

\begin{table}
  \centering
  \caption{\bf Data collected from Twitter related to the HUS/EHEC outbreak in Germany during May and June, 2011.}
  \label{tab:dataCollectionStats}
  \vspace{4pt}
  \begin{tabular}{ p{0.7\columnwidth}  r } 
   \hline
	\textbf{Description} & \textbf{Amount} \\ 
	\hline \hline
	Number of tweets collected related to medical conditions during May and June, 2011 &  7,710,231 \\
	\hline
	Tweets extracted related to the EHEC/HUS outbreak out of the ones collected & 456,226  \\
	\hline
	Distinct users that produced the tweets related to the outbreak & 54,381 	   			  \\
	\hline
\end{tabular}
\vspace{-10pt}
\end{table}

\subsection{Detection Methods}
The surveillance algorithms we used are well documented in the disease aberration literature e.g. \cite{DBLP:journals/jbi/Khan07,Hutwagner_Thompson_Seeman_Treadwell_2003,BassevilleNikiforov1993}. The objective of these algorithms is to detect aberration patterns in time series data when the volume of an observation variable exceeds an expected threshold value. In our case, for example, the observation variable corresponds to mentions of medical condition ``EHEC" withing the tweets.

The five biosurveillance algorithms we used for early detection are: the Early Aberration Reporting System (EARS) (1) C1, (2) C2, and (3) C3 algorithms, (4) F-statistic, and (5) Exponential Weighted Moving Average (EWMA). Please refer to~\cite{DBLP:journals/jbi/Khan07} for a detailed introduction.

We signal an alarm if the test statistic reported by the detection methods exceeds a threshold value, which is determined experimentally. The larger the amount by which the threshold is exceeded, the greater the severity of the alarm. 

Table~\ref{tab:detectionMethods} summarizes the alarm dates and detection methods parametrization, which follows the guidelines of N. Collier~\cite{2010_collier}. 

\begin{table}[t!]
\centering
\caption{\bf Detection method parameters and alarm dates}
\vspace{4pt}
\begin{tabular}{p{0.17\columnwidth} p{0.5\columnwidth} p{0.18\columnwidth}} 
\hline
\textbf{Detection Method} & \textbf{Parametrization}\;\cite{DBLP:journals/jbi/Khan07,2010_collier} & \textbf{Alarm Dates} \\
\hline \hline 
C1 & Training window = 15 days; buffer = 5 days;  upper control limit = $\mu + 3 \sigma$ & May 20 to May 28\\
\hline
C2 & Training window = 15 days; buffer = 5 days; upper control limit = $\mu + 3 \sigma$; alarm threshold=0.2 & May 20 to May 28\\
\hline
C3 & Training window = 15 days; buffer = 5 days; upper control limit = $\mu + 3 \sigma$; alarm threshold=0.3 & May 20 to May 24\\
\hline
F-statistic & Training window =15 days; buffer = 5 days; alarm threshold=0.6 & May 20 to June 30\\
\hline
EWMA & Training window =15 days; buffer = 5 days; alarm threshold=4, $\omega=0.2$4 & May 20 to May 30\\
\hline
\end{tabular}
\vspace{-10pt}
\label{tab:detectionMethods}
\end{table}

Using any of the detection methods (Table~\ref{tab:detectionMethods}), a daily count less than five tweets was enough to signal an alert on May 20th, 2011. The Early Warning and Response System (EWRS)~\footnote{\textbf{EWRS}: ewrs.ecdc.europa.eu} of the European Union received a first communication by the German authorities on Sunday May 22. MedISys~\footnote{\textbf{MedISys}: medusa.jrc.it/medisys} detected the first media report in the German newspaper \textit{Die Welt}~\footnote{\textbf{Die Welt}: welt.de} on Saturday May 21~\cite{Linge2011} and ProMED-mail~\footnote{\textbf{ProMED-mail}: promedmail.org} and all other major early alerting systems (e.g., ARGUS, Biocaster, GPHIN, HealthMap, PULS) covered the event on Monday May 23. 

Why was this early detection possible with respect to well established early warning systems? We tracked only Twitter as source of information, in contrast to MedISys for example, that tracks hundreds of news sources on the Internet. We consider Twitter's \textit{diversity} was the key element that helped in the earlier detection of the event. 

Twitter is a diverse stream of \textit{multiple sources}. In Twitter converges the \textit{contribution from the crowd} - millions of individual users obscure and renown; big and small media outlets; global and local newspapers, etc. Our work and that of MedISys focus on an analysis at a national level, but there are cases where support for the \textit{local perspective} is important, for example local and smaller news papers reaching a broader audience through Twitter. 

\begin{figure}[h!]
\centering
\includegraphics[width=1\columnwidth]{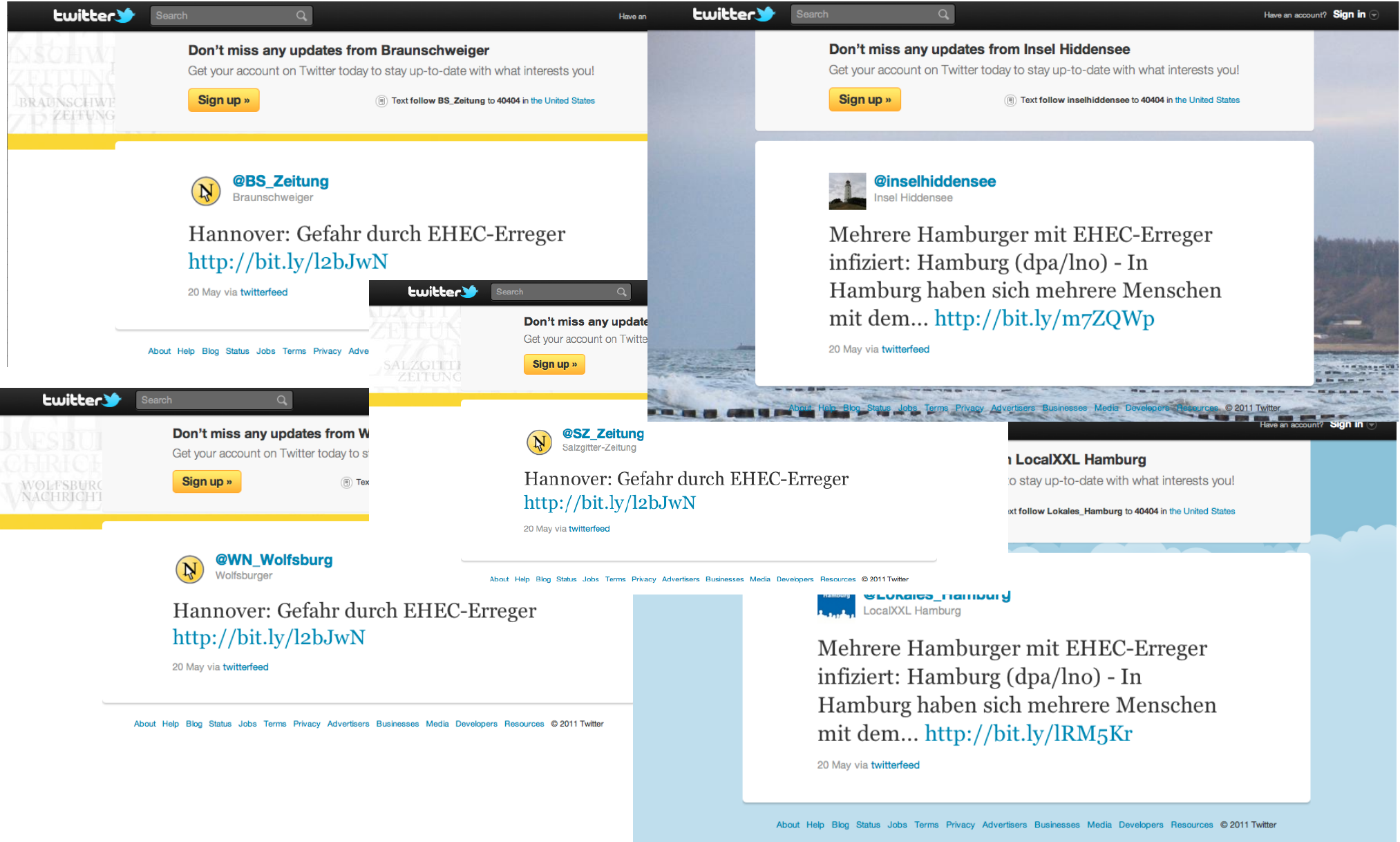}
\caption{\bf The 5 tweets that triggered the first signal for disease \textit{EHEC} on May 20, 2011.}
\label{fig:EHEC_early_tweets}
\vspace{-10pt}
\end{figure}

A closer look to day May 20, reveals that the first alarm was triggered based on five tweets, the actual messages are shown in Figure \ref{fig:EHEC_early_tweets}, all of them generated from sources not far from where the first cases of the outbreak were reported. Those users acted as \textit{local sensors}, producing tweets that spread the news faster than major newspapers.

\section{Twitter for Outbreak Analysis and Control}
\label{sec:analysisAndControl}
For public health officials, who are participating in the investigation of an outbreak, the millions of documents produced over social media streams represent an overwhelming amount of information for risk assessment. 

To reduce this overload we explore to what extent recommender systems techniques can help to filter information items according to the public health users' context and preferences (e.g., disease, symptoms, location). In particular, we focus on a personalized learning to rank approach that ultimately offers the user the most relevant and attractive tweets for risk assessment. In this section, we introduce our approach and report an experimental evaluation on the EHEC/HUS dataset collected from Twitter.

\subsection{Background: Learning to Rank for IR} 
\label{sec:L2R_intro}
Learning to rank for Information Retrieval (L2R) is an active area of recent research \cite{letor}. L2R is set as a supervised learning task that considers
fundamentally two phases: learning and retrieval. In learning (training), a
collection of queries and their corresponding retrieved documents are given.
Furthermore, the labels (i.e., relevance judgments) of the document with respect
to the queries are also available. The relevance judgments, provided by human
annotators, can represent ranks (e.g., categories in a total order) or binary labels (e.g., relevant or not-relevant). The
objective of learning is to construct a ranking model $\vec{w}$, e.g., a ranking
function, that achieves the best result on test data in the sense of optimization of a
performance measure (e.g., error rate, degree of agreement between the two
rankings, classification accuracy or mean average precision).

In retrieval (test phase), given a query-document pair, the learned ranking function  is
applied, returning a ranked  list of documents in descending order of their
relevance scores. More formally, suppose that $Q=\{q_1,\cdots,q_{|Q|}\}$ is the set of queries,
and $D=\{d_{1},\cdots,d_{|D|}\}$ the set of documents, the training set is
created as a set of query-document pairs, $(q_i, d_j) \in Q \times D $, upon
which a relevance judgment (e.g., a label) indicating the relationship between
$q_i$ and $d_j$ is assigned by an annotator. Suppose that
$Y=\{y_{1},\cdots,y_{|Y|}\}$ is the set of labels and $y_{ij} \in Y$ denotes the
label of query-document pair $(q_i, d_j)$. A feature vector $\phi(q_i, d_j)$ is
created from each query-document pair $(q_i, d_j), i=1,2,\cdots,|Q|;
j=1,2,\cdots,|D|$. The training set is denoted as $T=\{(q_i,d_j), \phi(q_i, d_j),
y_{ij}\}$. The ranking model is a real valued function of features:
\begin{equation}
   f(q, d) = \vec{w}\cdot\phi(q, d)
 \label{eq:rankingFunction}
 \end{equation}
where $\vec{w}$ denotes a weight vector. In ranking, for query $q_i$ the model
associates a score to each of the documents $d_j$ as their degree of relevance
with respect to query $q_i$ using $f(q_i,d_j)$, and sort the documents based on
their scores.

Table~\ref{tab:notation} gives a summary of notations described above.

\begin{table}[t!]
\centering
\caption{\bf Learning to Rank: Summary of Notations}
\vspace{2pt}
{\small
\begin{tabular}{l l} \hline
\textbf{Notations} & \textbf{Explanations}\\ \hline \hline 
$Q=\{q_1,\cdots,q_{|Q|}\}$ & Set of queries\\ 
$q_i \in Q$ & Query\\ 
$D=\{d_{1},\cdots,d_{|D|}\}$ & Set of documents\\ 
$d_j \in D$ & Document\\ 
$Y=\{y_{1},\cdots,y_{|Y|}\}$ & Set of relevance judgments\\ 
$y_{ij} \in Y$ & Relevance judgment of \\ 
               & query-document pair\\ 
               & $(q_i, d_j) \in Q \times D $\\ 
$\phi(q_i, d_j)$ & Feature vector w.r.t. $(q_i, d_j)$ \\
$\phi_k(q_i, d_j)$ & $k^{th}$ dimension of $\phi(q_i, d_j)$ \\
$T=\{(q_i,d_j), \phi(q_i, d_j), y_{ij}\}$ & Training set \\
$\substack{1 \leq i \leq |Q| \\ 1 \leq j \leq |D|}$ &  \\
\hline\end{tabular}
}
\label{tab:notation}
\end{table}

\textit{Pairwise} approaches, such as Ranking SVM~\cite{Joachims:2002:OSE:775047.775067} or Stochastic Pairwise Descent~\cite{2009_scully}, have proved successful in addressing the L2R task. A comprehensive study on different learning to rank techniques can be found in~\cite{Liu:2009:LRI:1618303.1618304}.

Although much work has been carried out on L2R techniques for systems that return documents in response to a query, little effort has been devoted to exploiting L2R in a personalized setting, specially in the domain of epidemic intelligence.

\subsection{Our Approach:\\Ranking Tweets for Epidemic Intelligence} 
We propose to use the user \textit{context} as implicit criteria to select tweets of potential relevance, that is, we will rank and derive a short list of tweets based on the user context. The user context $C_u$ is defined as a triple 
\begin{equation}
\label{eq:userContext}
C_u = (t, MC_u, L_u) \;\;,
\end{equation}
where $t$ is a discrete time interval, $MC_u$ the set of medical conditions, and  $L_u$ the set of locations of user interest. 

We define three concepts that will help us to discuss our approach in rest of the section:

\textbf{Medical Condition} is a string that describes a human medical condition, such as a disease, disorder or syndrome. We represent the set of medical conditions as $MC$.

\textbf{Location} is a string that is used to identify a point or an area on the Earth's surface, which can be mapped to 
a specific pairing of latitude and longitude. The set of locations is denoted as $L$.

\textbf{Complementary Context} is defined as the set of nouns, which are neither Locations nor Medical Conditions.  Complementary Context may include named entities such as names of persons, organizations, affected organisms, expressions of time, quantities, etc. We denote the set of named entities that represents the complementary context as  $CC$, where $CC \cap (L \cup MC) = \emptyset$.

\begin{algorithm}[t!]
\renewcommand{\algorithmicrequire}{\textbf{Input:}}
\renewcommand{\algorithmicensure}{\textbf{Output:}}
\caption{Personalized Tweet Ranking algorithm for Epidemic Intelligence (PTR4EI)}
\label{algo:tweetRank}
\begin{algorithmic}[1]
\REQUIRE User Context $C_u = (t, MC_u, L_u)$, \\Inverted index $\mathcal{T}$ of tweets collected for epidemic intelligence before time $t$
\ENSURE Ranking Function $f_{C_u}$ for User Context $C_u$
\vspace{5pt}
\STATE Compute LDA topics ($topicsLDA$) on $\mathcal{T}$
\vspace{5pt}
\STATE Consider each $mc \in MC_u$ as a hash-tag, and extract from $\mathcal{T}$ all co-occurring hash-tags: $coHashTags$
\vspace{5pt}
\STATE Classify the terms in $topicsLDA$ and the hash-tags in $coHashTags$ as Medical Condition $MC_x$, Location $L_x$ or Complementary Context $CC_x$
\vspace{5pt}
\STATE Build a set of queries as follows: \\
$Q = \{ q \;|\; q \in MC_u \times \mathcal{P}(\{L_u \cup  MC_x \cup L_x \cup CC_x\})\}$
\vspace{5pt}
\STATE For each query $q_i \in Q$ obtain tweets $D$ from the collection $\mathcal{T}$
\vspace{5pt}
\STATE Elicit relevance judgments $Y$ on a subset $D_y \subset D$
\vspace{5pt}
\STATE For each tweet $d_j \in D$, obtain the feature vector $\phi(q_i, d_j)$ w.r.t. $(q_i, d_j) \in Q \times D$
\vspace{5pt}
\STATE Apply learning to rank to obtain a ranking function for the user context $C_u$:
$f_{C_u}(q, d) = \vec{w}\cdot\phi(q, d)$
\vspace{5pt}
\RETURN $f_{C_u}(q, d)$
\end{algorithmic}
\end{algorithm}

Out Personalized Tweet Ranking for Epidemic Intelligence algorithm or PTR4EI is shown in Algorithm~\ref{algo:tweetRank}. The algorithm extends a learning to rank framework (Section ~\ref{sec:L2R_intro} by considering a personalized setting that exploits user's individual context. 

More precisely, we consider the context of the user, $C_u$, and prepare a set of queries, $Q$, for a target event (e.g., a disease outbreak). We first compute LDA~\cite{LDA} on an indexed collection $\mathcal{T}$ of tweets for epidemic intelligence, where not all tweets are necessarily interesting for the target event. 

We also extract the hash-tags that co-occur with the user context by considering the medical conditions and locations in $C_u$ as hash-tags themselves, and find which other hash-tags co-occur with them within a tweet, and how often they co-occur, which will help us to select the most representative hash-tags for the target event. 

The set $Q$ is constructed by expanding the original terms in $C_u$ with the ones in the LDA topics and co-occurring hash-tags, which are previously classified as medical condition, location or complementary context. 

We build the set $D$ of tweets by querying index $\mathcal{T}$ using $q \in Q$ as query terms. Next, we elicit judgments from experts on a subset of the tweets retrieved, in order to construct $D_y \subset D$. 

We then obtain for each tweet $d_j \in D$ its features vector $\phi(q_i, d_j)$ with respect to the pair $(q_i, d_j) \in Q \times D$. 

Finally and with these elements, we apply a learning to rank algorithm to obtain the ranking function for the given user context. 

In the rest of the section, we evaluate our approach considering as event of interest the EHEC/HUS outbreak in Germany, 2011.

\subsubsection{Experiments and Evaluation}
To support users in the assessment and analysis during the EHEC/HUS outbreak, we set the user context (Eq.~\ref{eq:userContext}) as $C_u = (t, MC_u, L_u) = ([\text{2011-05-23; 2011-06-19}], \{\text{``EHEC"}\}, \{\text{``Lower Saxony"}\})$, in this way, we are taking into account the main period of the outbreak~\footnote{Please note, that even though the main period of the outbreak is considered for the evaluation, nothing prevents us to build the model during the ongoing outbreak, and recompute it periodically (e.g., weekly).}, the disease of interest, and the German state with more cases reported.

Following Algorithm~\ref{algo:tweetRank}, we computed LDA and extracted the co-occurring hash-tags using the indexed collection $\mathcal{T}$ described in Section \ref{sec:dataCollection}. Table~\ref{tab:EHEC_LDA_Topics} shows four LDA topics for each week of the time period of interest, and Table~\ref{tab:EHEC_cooccurringHashtags} presents the hash-tags co-occurring with \textit{\#EHEC}.

\begin{table}[t!]
  \centering
  \caption{\bf Four LDA topics (columns) computed weekly during the main period of the outbreak: from May 23 to June 19, 2011. We classify terms within each topic as \textit{Medical Condition (MC)}, \textit{Location (L)}, or \textit{Complementary Context (CC)}. 
%Terms outside these categories are ignored.
  }
  \label{tab:EHEC_LDA_Topics}
  \scalebox{0.63}{
  \begin{tabular}{| l | l | l | l |} 
\hline
\multicolumn{4}{| c |}{\textbf{Week 21}}\\
\hline
EHEC \textit{(MC)}  & fever \textit{(MC)}  & EHEC \textit{(MC)}  & EHEC \textit{(MC)}  \\
cucumbers \textit{(CC)} & pain \textit{(MC)}  & casualty  \textit{(-)} & pathogen \textit{(MC)}  \\
Spain \textit{(L)}  & headache \textit{(MC)} & women \textit{(CC)} & Northern Germany \textit{(L)} \\
tomatoes \textit{(CC)} & sniff \textit{(MC)} & intestinal germ \textit{(MC)} & diarrhea \textit{(MC)} \\
salad \textit{(CC)} & pain \textit{(MC)} & panic \textit{(MC)} & dead \textit{(MC)} \\
\hline
\multicolumn{4}{|c|}{\textbf{Week 22}}\\
\hline
EHEC \textit{(MC)}	&	EHEC \textit{(MC)}	&	EHEC \textit{(MC)}	&	EHEC \textit{(MC)}	\\
dead \textit{(MC)}	&	intestinal germ \textit{(MC)}	&	cucumbers \textit{(CC)} &	cucumbers \textit{(CC)}\\
Germany \textit{(L)} &	source	\textit{(-)}  &	pathogen \textit{(MC)}	&	salad \textit{(CC)}\\
people	\textit{(-)}  &	search	\textit{(-)}  &	Spain \textit{(L)} &	pain \textit{(MC)}	\\
live \textit{(-)}  &	Hamburg \textit{(L)} &	farmers \textit{(CC)} &	women	\textit{(CC)}\\
\hline
\multicolumn{4}{|c|}{\textbf{Week 23}}\\
\hline
EHEC \textit{(MC)}	&	headache	 \textit{(MC)}&	EHEC \textit{(MC)} &	 EHEC \textit{(MC)}\\
cucumber	\textit{(CC)} &	pain \textit{(MC)}	&	cucumbers \textit{(CC)} &	sprout \textit{(CC)}\\
eu	 \textit{(CC)} &	fever \textit{(MC)} &	sprout \textit{(CC)} &	source	\textit{(-)} \\
crisis management	\textit{(-)}  &	people	\textit{(-)}  &	pathogen \textit{(MC)}	&	suspicion	\textit{(-)} \\
farmers \textit{(CC)} &	cough \textit{(MC)} &	salad \textit{(CC)} &	hus \textit{(MC)}\\
\hline
\multicolumn{4}{|c|}{\textbf{Week 24}}\\
\hline
EHEC \textit{(MC)} &	headache \textit{(MC)} &	stomach ache \textit{(MC)}&	pain \textit{(MC)}\\
germ \textit{(MC)} &	fever	 \textit{(MC)} &	sniff \textit{(MC)} &	bellyache \textit{(MC)}\\
sprout \textit{(CC)}& slept \textit{(-)}  &	pain \textit{(MC)} &	cough \textit{(MC)}\\
health \textit{(MC)} & sniff \textit{(MC)} &	regions \textit{(-)}  &	throat \textit{(CC)}\\
all-clear \textit{(CC)} &	head	 \textit{(CC)} &	examined	\textit{(-)} &	sniff \textit{(MC)}\\
\hline
\end{tabular}
}
\vspace{-10pt}
\end{table}

\begin{table}[t!]
    \caption{\bf Hash-tags co-occurring with \textit{\#EHEC} during May 23 and June 19, 2011, the main period of the outbreak. The hash-tags are classified as entities of type \textit{Medical Condition}, \textit{Location}, or \textit{Complementary Context}, hash-tags out of these categories are discarded.\\}
  \label{tab:EHEC_cooccurringHashtags}
  \scalebox{0.8}{
  \begin{tabular}{| c | c | c  c |} 
\hline
\textbf{Medical Condition} & \textbf{Location} & \multicolumn{2}{| c |}{\textbf{Complementary Context}} \\	
\hline
\hline
\multicolumn{4}{| c |}{\textbf{Week 21}}\\						
\hline			
bacteria & bremen & cucumber\_salad & cdu \\
diarrhea & cuxhaven & cucumbers & edeka \\
ehec\_victim & hamburg &  ehec\_vegetable & fdp \\
hus & m\"{u}nster &  tomatoes & merkel \\
intestinal\_infection & northern\_germany & vegetables & rki \\
\hline
\multicolumn{4}{| c |}{\textbf{Week 22}}\\						
\hline
bacteria & berlin & cucumbers & bild \\
diarrhea & germany & obst & fdp \\
ehec\_pathogen & hamburg & salad & n24 \\
hus & l\"{u}beck & terror & rki \\
intestinal\_infection & spain & tomatoes & rtl \\
\hline
\multicolumn{4}{| c |}{\textbf{Week 23}}\\						
\hline
bacteria & bavaria & cucumbers & ehec\_freei \\
diarrhea & berlin & salad & fdp \\
ehec\_pathogen & germany & sojasprout & merkel \\
hus & hamburg & sprout & n24 \\
intestinal\_infection & lower\_saxony &   & rki \\
\hline
\multicolumn{4}{| c |}{\textbf{Week 24}}\\						
\hline
bacteria & lower\_saxony & donate\_blood &  \\
died & &  ehec\_free &  \\
health &  &sojasprout &  \\
hus &  &  &  \\
\hline							
\end{tabular}
}
\vspace{-10pt}
\end{table}

We asked three experts: one from the Robert Koch Institute and the other two from the Lower Saxony State Health Department (NLGA)~\footnote{\textbf{NLGA}: nlga.niedersachsen.de} to provide their individual judgment on a subset $D_y$ of 240 tweets, evaluating for each tweet, if it was relevant or not to support their analysis of the outbreak. Any disagreement in the assigned relevance scores were resolved by majority voting.

We selected these tweets from the index $\mathcal{T}$ as follows: 30 were obtained using as query the term ``EHEC", i.e., $MC_u$, together with the medical conditions identified using LDA, and 30 using the medical conditions from the hash-tags. We used a similar procedure combining query ``EHEC" with the locations and complementary context extracted from LDA and hash-tag co-occurrence, obtaining 30 tweets at every step, for a total of 120 tweets. For the rest 60, we used  the query term ``EHEC" alone, then we ordered the result set chronologically based on the tweets' publication date, and selected the most recent ones. 

We prepared five binary features for each tweet as follows: 
\vspace{4pt}

\begin{tabular}{l p{0.7\columnwidth}} \hline
Feature &  Value = True\\ \hline \hline 
 $F_{MC}$ & If a medical condition is present in the tweet \\
 $F_{L}$ &  If a location is present in the tweet\\
 $F_{\text{\#-tag}}$ & If a hash-tag is present in the tweet \\
 $F_{CC}$ &  If a complementary context term is present in the tweet \\
 $F_{URL}$ & If a URL is present in the tweet\\
\hline
\end{tabular}
\vspace{4pt}

For learning the ranking function, we used Stochastic Pairwise Descent (SPD) algorithm~\cite{2009_scully}, which solves the same optimization problem as Ranking SVM~\cite{Joachims:2002:OSE:775047.775067}, but using stochastic gradient descent, whose characteristics make it more appealing to scale to larger datasets (e.g.,~\cite{bottou-2010}).

We compared our approach, that expand the user context with latent topics and social generated hash-tags, against two ranking methods:

\begin{itemize}
    \item \textbf{RankMC}: It learns a ranking function using only medical conditions as feature, i.e., $F_{MC}$. Please note, that this baseline also considers related medical conditions to the ones in $MC_u$, which makes it stronger than non-learning approaches, such as BM25 or TF-IDF scores, that use only the $MC_u$ elements as query terms.
    
    \item \textbf{RankMCL}: It is similar to RankMC, but besides the medical conditions, it uses a local context to perform the ranking (i.e., features: $F_{MC}$ and $F_{L}$). We expect this method to perform better than RankMC, since it does not only take into account the spatial information from the user context, but also additional locations in the collection.
\end{itemize}

We conducted 10-fold cross validation experiments. For each fold, we used 80\%  of the tweets for training and the remaining 20\% for testing. The test set is used to evaluate the ranking methods. The reported performance is the average over the ten folds.

\subsubsection*{Evaluation Measures}
\label{sec:evalMeasures}
For evaluation, we used three evaluation measures widely used in information
retrieval, namely precision at position $n$ ($P@n$), mean average precision
(MAP), and normalized discount cumulative gain (NDCG). Their definitions are as
follows.

\textbf{Precision at Position $n$ ($P@n$)}~\cite{2011_ricBaeza} measures the relevance of the top $n$ 
documents in the ranking list with respect to a given query:

 \begin{equation}
	P@n = \frac{\text{\# of relevant docs in top $n$ results}}{n}
 \label{eq:patn}
 \end{equation}

\textbf{Mean Average Precision (MAP)}
\label{sec:map}
The average precision (AP)~\cite{2011_ricBaeza} of a given query is calculated as
Eq.~(\ref{eq:ap}), and corresponds to the average of $P@n$ values for all relevant documents:

 \begin{equation}
	AP = \frac{\sum_{n=1}^{N}{(P@n * rel(n))}}{\text{\# of relevant docs for
	this query}}
 \label{eq:ap}
 \end{equation}

where $N$ is the number of retrieved documents, and $rel(n)$ is a binary
function that evaluates to 1 if the $n^{th}$ document is $relevant$, and 0
otherwise. Finally, MAP~\cite{2011_ricBaeza} is obtained averaging the AP
values over the set of queries.

\textbf{Normalized Discount Cumulative Gain (NDCG)}
For a single query, the NDCG~\cite{ndcg} value of its ranking list at position $n$
is computed by Eq.~(\ref{eq:ndcg}):

\begin{equation}
	NDCG@n = Z_n\sum_{j=1}^n{\frac{2^{r(j)}-1}{log(1+j)}}
 \label{eq:ndcg}
 \end{equation}

where $r(j)$ is the rating of the the $j$-th document in the ranking list, and
the normalization constant $Z_n$ is chosen so that the perfect list gets NDCG
score of 1. 

For the training dataset, we define two ratings $\{1, 0\}$
corresponding to ``\emph{relevant to the outbreak}'' and ``\emph{not-relevant to the outbreak}'' in order to compute NDCG scores.

\subsubsection*{Results}
\begin{figure}[t!]
\centering
\includegraphics[width=1\columnwidth]{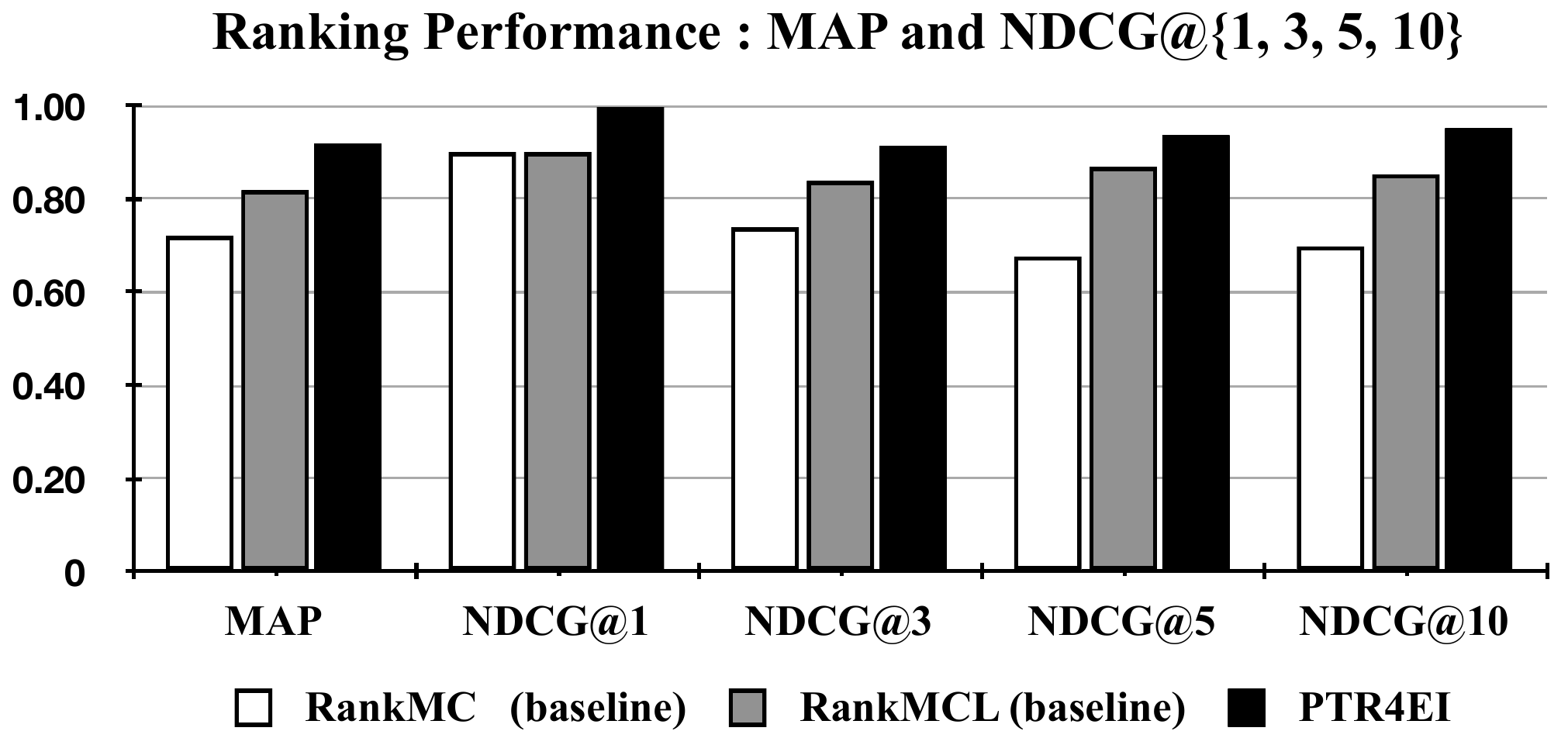}
\caption{\bf MAP and NDCG Results}
\label{fig:results_MAP_NDCG}
\end{figure}

The ranking performance in terms of precision is presented in Table~\ref{tab:resultsPrecision}, MAP and NDCG results are shown in Figure~\ref{fig:results_MAP_NDCG}.  As we can appreciate PTR4EI outperforms both baselines. Local information helps RankMCL to beat RankMC, for example MAP improves from 71.96\% (RankMC) up to 81.82\% (RankMCL). PTR4EI, besides local features, exploits complementary context information and particular Twitter features, such as the presence of hash-tags or URLs in the tweets, this information allows it to improve its ranking performance even further, reaching a MAP of 91.80\%. A similar behavior is observed for precision and NDCG, where PTR4EI is statistically significantly better than RankMC and RankMCL.

\begin{table}[t!]
\centering
\caption{\bf \small{Ranking Performance in terms of P@\{1, 3, 5, 10\}}}
\vspace{4pt}
\scalebox{0.95}{
\begin{tabular}{c c c c c} \hline
\textbf{Method} & \textbf{P@1} & \textbf{P@3} & \textbf{P@5} & \textbf{P@10} \\
\hline \hline 
RankMC   (baseline)	 & 90~\%  & 73.34~\% & 64~\% & 69~\% \\
RankMCL (baseline)	 & 90~\%  & 83.33~\% & 88~\% & 85~\% \\
PTR4EI	                    & \textbf{100~\%} & \textbf{90~\%} & \textbf{94~\%} & \textbf{96~\%} \\
\hline
\end{tabular}
}
\label{tab:resultsPrecision}
\end{table}

%\section{Related Work}
\section{Related Work}
\label{sec:relwork}
%% Detection
In order to detect public health events, supervised \cite{Stewart:2011}, unsupervised \cite{DBLP:conf/spire/FisichellaSCD11} and rule-based approaches have been used to extract public health events from social media and news.  For example, PULS \cite{SteFua+08}  identify the disease, time, location and cases of a news-reported event. It is integrated into MedISys, which automatically collects news articles concerning public health in various languages, and aggregates the extracted facts according to pre-defined categories, in a multi-lingual manner.

%% Correlation and Content Analysis
Other systems have sought to use the web and social media as a predictor to monitor and gauge the seasonal patterns of influenza. These systems correlate the queries used in search behavior with the infection rates of influenza-like illnesses statistics~\cite{Polgreen_usinginternet,Ginsberg2009}.

Monitoring analysis has also been carried out on Twitter. The work of Chew et al. focused on the use of the terms ``H1N1" and ``swine flu"  during the H1N1 2009 outbreak \cite{chew-2009}. They showed that the concise  and timely nature of tweets can provide health officials with the a means to become aware, and respond to concerns raised by the public. 

Culotta applied text classification to filter out tweets that are not reporting about influenza-like illnesses. Further, they modeled influenza rates by regression models and compared to U.S. Center of Disease Control statistics \cite{Culotta:2010}.

Lampos and Cristianini also presented a monitoring tool for social media that is based on the textual analysis of micro-blog content \cite{Lampos2010}, \cite{Lampos2010b}.  Their study focused on influenza-like illnesses in the UK and showed a correlation with data from the Health Protection Agency. Another study of Twitter content concentrated on influenza-like illnesses in the U.S. \cite{Signorini2011}. Paul and Dredze \cite{Paul_Dredze_2011,DBLP:conf/icwsm/PaulD11} introduced a new aspect topic model for Twitter that associates symptoms, treatments and general words with diseases. Their focus is on general public health, not necessarily infectious diseases or disease outbreaks.

In contrast to these systems, we seek to not only detect and monitor potential public health threats, but also provide support for public health officials to asses the potential risk associated with the volume of information that is available within Twitter streams. Moreover, our  proposed approach shows the  potential of using Twitter for monitoring non-seasonal outbreaks in and geo-spacially sparse tweet locations.

Our work is similar to that of ~\cite{Linge2011}, were media reports on the 2011 EHEC outbreak in Germany are tracked. Although in their work no early warning was possible, they identified key aspects of  developing outbreak stories. In contrast to this work, our approach exploits social media data and we show that a system can help to get early warnings on public health threats. 
%Ranking

Although some works exist that address the task of ranking tweets, little effort has been devoted to explore personalized ranking of tweets in the domain of epidemic intelligence. For example Duan et al.  rank individual generic tweets according to their relevance to a given query \cite{duan2010}. The features used include content relevance features, Twitter specific features and  account authority features. In contrast, our is a personalized learning to rank approach for epidemic intelligence, that exploits an expanded user context by means of latent topics and on social hash-tagging behavior.

\section{Conclusion and Future Directions}
\label{sec:conclusionAndFutureDirection}
To show the potential of Twitter for early warning, we focused on the recent EHEC/HUS outbreak in Germany, and monitor the social stream. We applied several  biosurveillance methods on a set of tweets collected in real time during the time of the event using Twitter API. All the detection methods triggered an alarm on May 20, a day ahead of well established early warning systems, such as MedISys. 

After the detection of the outbreak, authorities investigating the cause and the impact in the population were interested in the analysis of micro-blog data related to the event. Thousands of tweets were produced every day, which made this task overwhelming for the experts. We proposed in this work a Personalized Tweet Ranking algorithm for Epidemic Intelligence (PTR4EI) that provides users a personalized short list of tweets that meets the context of their investigation. PTR4EI exploits features that go beyond the medical condition and location (i.e., user context), but includes complementary context information, extracted using LDA and the social hash-tagging behavior in Twitter, plus additional Twitter specific features. Our experimental evaluation showed the superior ranking performance of PTR4EI.

We are currently working closely with  German and global public health institutions to help them integrate the monitoring of social media to their existing surveillance systems. 

As future work, we plan to scale up our experiments, and to apply techniques of online ranking in order to update the model more efficiently as the outbreak develops.

We have shown the potential of Twitter to trigger early warnings in the case of sudden outbreaks and how personalized ranking for epidemic intelligence can be achieved. We believe our work can serve as a building block for an open early warning system based on Twitter, and hope that this paper provides some insights into the future of epidemic intelligence based on social media streams.
%\section{Acknowledgments}
%MECO Team
\begin{small}
\bibliographystyle{aaai} 
\bibliography{biblio}
\end{small}
\flushend
\end{document}